\documentstyle[twoside,psfig]{article}

\catcode`\@=11
\long\def\@makefntext#1{
\protect\noindent \hbox to 3.2pt {\hskip-.9pt  
$^{{\eightrm\@thefnmark}}$\hfil}#1\hfill}		

\def\@makefnmark{\hbox to 0pt{$^{\@thefnmark}$\hss}}	
	
\def\ps@myheadings{\let\@mkboth\@gobbletwo
\def\@oddhead{\hbox{}
\rightmark\hfil\eightrm\thepage}   
\def\@oddfoot{}\def\@evenhead{\eightrm\thepage\hfil
\leftmark\hbox{}}\def\@evenfoot{}
\def\sectionmark##1{}\def\subsectionmark##1{}}

\oddsidemargin=\evensidemargin
\newcounter{sectionc}\newcounter{subsectionc}\newcounter{subsubsectionc}
\renewcommand{\section}[1] {\vspace{12pt}\addtocounter{sectionc}{1} 
\setcounter{subsectionc}{0}\setcounter{subsubsectionc}{0}\noindent 
	{\tenbf\thesectionc. #1}\par\vspace{5pt}}
\renewcommand{\subsection}[1] {\vspace{12pt}\addtocounter{subsectionc}{1} 
	\setcounter{subsubsectionc}{0}\noindent 
	{\bf\thesectionc.\thesubsectionc. {\kern1pt \bfit #1}}\par\vspace{5pt}}
\renewcommand{\subsubsection}[1] {\vspace{12pt}\addtocounter{subsubsectionc}{1}
	\noindent{\tenrm\thesectionc.\thesubsectionc.\thesubsubsectionc.
	{\kern1pt \tenit #1}}\par\vspace{5pt}}

\newcounter{appendixc}
\newcounter{subappendixc}[appendixc]
\newcounter{subsubappendixc}[subappendixc]
\renewcommand{\thesubappendixc}{\Alph{appendixc}.\arabic{subappendixc}}
\renewcommand{\thesubsubappendixc}
	{\Alph{appendixc}.\arabic{subappendixc}.\arabic{subsubappendixc}}

\renewcommand{\appendix}[1] {\vspace{12pt}
        \refstepcounter{appendixc}
        \setcounter{figure}{0}
        \setcounter{table}{0}
        \setcounter{lemma}{0}
        \setcounter{theorem}{0}
        \setcounter{corollary}{0}
        \setcounter{definition}{0}
        \setcounter{equation}{0}
        \renewcommand{\thefigure}{\Alph{appendixc}.\arabic{figure}}
        \renewcommand{\thetable}{\Alph{appendixc}.\arabic{table}}
        \renewcommand{\theappendixc}{\Alph{appendixc}}
        \renewcommand{\thelemma}{\Alph{appendixc}.\arabic{lemma}}
        \renewcommand{\thetheorem}{\Alph{appendixc}.\arabic{theorem}}
        \renewcommand{\thedefinition}{\Alph{appendixc}.\arabic{definition}}
        \renewcommand{\thecorollary}{\Alph{appendixc}.\arabic{corollary}}
        \renewcommand{\theequation}{\Alph{appendixc}.\arabic{equation}}
        \noindent{\tenbf Appendix \theappendixc #1}\par\vspace{5pt}}
\newcommand{\subappendix}[1] {\vspace{12pt}
        \refstepcounter{subappendixc}
        \noindent{\bf Appendix \thesubappendixc. {\kern1pt \bfit #1}}
	\par\vspace{5pt}}
\newcommand{\subsubappendix}[1] {\vspace{12pt}
        \refstepcounter{subsubappendixc}
        \noindent{\rm Appendix \thesubsubappendixc. {\kern1pt \tenit #1}}
	\par\vspace{5pt}}

\newcommand{\textlineskip}{\baselineskip=13pt}
\newcommand{\smalllineskip}{\baselineskip=10pt}

\def\eightcirc{
\begin{picture}(0,0)
\put(4.4,1.8){\circle{6.5}}
\end{picture}}
\def\eightcopyright{\eightcirc\kern2.7pt\hbox{\eightrm c}} 

\newcommand{\copyrightheading}[1]
	{\vspace*{-2.5cm}\smalllineskip{\flushleft
	{\footnotesize Modern Physics Letters A #1}\\
	{\footnotesize $\eightcopyright$\, World Scientific Publishing
	 Company}\\
	 }}

\newcommand{\publisher}[2]{{\begin{center}\footnotesize\smalllineskip 
	Received #1\\
	Revised #2
	\end{center}
	}}

\def\abstracts#1#2#3{{
	\centering{\begin{minipage}{4.5in}\footnotesize\baselineskip=10pt
	\parindent=0pt #1\par 
	\parindent=15pt #2\par
	\parindent=15pt #3
	\end{minipage}}\par}} 

\renewenvironment{thebibliography}[1]
	{\frenchspacing
	 \ninerm\baselineskip=11pt
	 \begin{list}{\arabic{enumi}.}
        {\usecounter{enumi}\setlength{\parsep}{0pt}     
	 \setlength{\leftmargin 12.7pt}{\rightmargin 0pt} 
         \setlength{\itemsep}{0pt} \settowidth
	{\labelwidth}{#1.}\sloppy}}{\end{list}}

\newcounter{itemlistc}
\newcounter{romanlistc}
\newcounter{alphlistc}
\newcounter{arabiclistc}

\newcommand{\fcaption}[1]{
        \refstepcounter{figure}
        \setbox\@tempboxa = \hbox{\footnotesize Fig.~\thefigure. #1}
        \ifdim \wd\@tempboxa > 5in
           {\begin{center}
        \parbox{5in}{\footnotesize\smalllineskip Fig.~\thefigure. #1}
            \end{center}}
        \else
             {\begin{center}
             {\footnotesize Fig.~\thefigure. #1}
              \end{center}}
        \fi}

\newcommand{\tcaption}[1]{
        \refstepcounter{table}
        \setbox\@tempboxa = \hbox{\footnotesize Table~\thetable. #1}
        \ifdim \wd\@tempboxa > 5in
           {\begin{center}
        \parbox{5in}{\footnotesize\smalllineskip Table~\thetable. #1}
            \end{center}}
        \else
             {\begin{center}
             {\footnotesize Table~\thetable. #1}
              \end{center}}
        \fi}

\def\@citex[#1]#2{\if@filesw\immediate\write\@auxout
	{\string\citation{#2}}\fi
\def\@citea{}\@cite{\@for\@citeb:=#2\do
	{\@citea\def\@citea{,}\@ifundefined
	{b@\@citeb}{{\bf ?}\@warning
	{Citation `\@citeb' on page \thepage \space undefined}}
	{\csname b@\@citeb\endcsname}}}{#1}}

\newif\if@cghi
\def\cite{\@cghitrue\@ifnextchar [{\@tempswatrue
	\@citex}{\@tempswafalse\@citex[]}}
\def\citelow{\@cghifalse\@ifnextchar [{\@tempswatrue
	\@citex}{\@tempswafalse\@citex[]}}
\def\@cite#1#2{{$\null^{#1}$\if@tempswa\typeout
	{IJCGA warning: optional citation argument 
	ignored: `#2'} \fi}}

\def\pmb#1{\setbox0=\hbox{#1}
	\kern-.025em\copy0\kern-\wd0
	\kern.05em\copy0\kern-\wd0
	\kern-.025em\raise.0433em\box0}

\def\fnt#1#2{\footnotetext{\kern-.3em
	{$^{\mbox{\scriptsize #1}}$}{#2}}}

\def\ps@myheadings{%
    \let\@oddfoot\@empty\let\@evenfoot\@empty
    \def\@evenhead{\slshape\leftmark\hfil}
    \def\@oddhead{\hfil{\slshape\rightmark}}
    \let\@mkboth\@gobbletwo
    \let\sectionmark\@gobble
    \let\subsectionmark\@gobble
    }
\font\tenrm=cmr10
\font\tenit=cmti10 
\font\tenbf=cmbx10
\font\bfit=cmbxti10 at 10pt
\font\ninerm=cmr9

\font\eightrm=cmr8

\def\qed{\hbox{${\vcenter{\vbox{			
   \hrule height 0.4pt\hbox{\vrule width 0.4pt height 6pt
   \kern5pt\vrule width 0.4pt}\hrule height 0.4pt}}}$}}

\def\Journal#1#2#3#4{{#1} {\bf #2}, #3 (#4)}
\def\FBS{\em Few-Body Syst.}

\def\LNC{\em Lett. Nuovo Cim.}

\def\NPA{{\em Nucl. Phys.} A}
\def\PLB{{\em Phys. Lett.}  B}
\def\PRL{\em Phys. Rev. Lett.}
\def\PRC{{\em Phys. Rev.} C}

\def\ZPA{{\em Z. Phys.} A}

\def\EPJA{{\em Eur. Phys. J.} A}
\def\JPG{{\em J. Phys.} G}

\begin{document}

\setlength{\textheight}{7.7truein}  

\thispagestyle{empty}

\markboth{\protect{\footnotesize\it Present Status of Electromagnetic 
Reactions on the Deuteron above Pion 
Threshold}}{\protect{\footnotesize\it Present 
Status of Electromagnetic Reactions on the Deuteron above Pion Threshold}}

\normalsize\textlineskip

\setcounter{page}{1}

\copyrightheading{}	

\hfill{\small {\bf MKPH-T-02-10}}\\
\vspace*{0.5truein}

\centerline{\bf PRESENT STATUS OF ELECTROMAGNETIC REACTIONS}
\baselineskip=13pt
\centerline{\bf ON THE DEUTERON ABOVE PION THRESHOLD\footnote{Supported by 
Deutsche Forschungsgemeinschaft (SFB 443).}}

\vspace*{0.4truein}
\centerline{\footnotesize H. ARENH\"OVEL, E.M. DARWISH\footnote{Supported 
by Deutscher Akademischer Austauschdienst. Present 
address: Physics Department, Faculty of Science, South Valley 
University, Sohag, Egypt.} , A. FIX, M. SCHWAMB}
\baselineskip=12pt
\centerline{\footnotesize\it Institut f\"ur Kernphysik, Johannes 
Gutenberg-Universit\"at}
\baselineskip=10pt
\centerline{\footnotesize\it 55099 Mainz, Germany}

\vspace*{0.228truein}

\publisher{(received date)}{(revised date)}

\vspace*{0.23truein}

\abstracts{
The present status of the theoretical description of electromagnetic 
reactions on the deuteron above pion threshold 
is reviewed. Three major topics are
considered: (i) retardation effects in $\pi$-meson exchange contributions to 
$NN$-interaction and meson exchange currents in deuteron photodisintegration 
above $\pi$-threshold in the $\Delta$-resonance region, (ii)
off-shell effects in the one-body current treated in a simple pion cloud 
model in deuteron photodisintegration in and above the $\Delta$-resonance 
region, and (iii) final state interaction effects in photoproduction of 
$\pi$ and $\eta$ mesons on the deuteron.
}{}{}

\section{Introduction}
\vspace*{-0.5pt}
\noindent
In view of the fact that at present QCD in the 
non-perturbative regime can at best be described 
by effective degrees of freedom only, 
one uses a framework with meson, nucleon and isobar
degrees of freedom. Hadron properties are either described 
phenomenologically or by effective quark models.
The central question then is: 
How accurate is this effective description, and where is the 
borderline beyond which explicit quark-gluon degrees of freedom have to 
be considered? It is very likely that no clear cut answer exists.

However, the study of electromagnetic reactions
on few-nucleon systems may give at least a partial answer because
lightest nuclei (deuteron, helium-3) allow reliable theoretical 
descriptions, and approximations, which are unavoidable in more complex 
many-body systems, are not necessary.
Therefore, such systems constitute reliable test laboratories 
for the investigation of effective degrees of freedom.
Furthermore, the electromagnetic interaction is well known and 
sufficiently weak, in order to allow conclusive 
interpretations in terms of charge and current matrix elements.
Finally, reactions above pion threshold are of particular interest 
with respect to explicit meson degrees of freedom and internal 
baryon structure.

The main points of interest are 
(i) the role of meson and isobar degrees of freedom in 
medium energy reactions,
(ii)
many-body phenomena, induced by the effective description 
in terms of meson, nucleon and isobar d.o.f., e.g., 
the role of pion retardation in $NN$ interaction and two-body 
meson exchange operators, and 
(iii)
properties of the neutron like, e.g., the elementary $\pi$ and 
$\eta$ photoproduction on the neutron, in other words, the use 
of light nuclei as effective neutron targets. 
Particularly suited are quasifree reactions on the deuteron 
in order to minimize final state interaction effects.
However, for a reliable interpretation it is mandatory to correct
for medium influences as, e.g., described by two-body effects.

\section{Pion Retardation in $\boldmath{d(\gamma,p)n}$ above Pion Threshold}
\vspace*{-0.5pt}
\noindent
Most sophisticated theories of photodisintegration in the 
$\Delta$ region, which are based on a coupled channel approach 
and use a $\Delta$ excitation operator from a fit of 
$\pi$ photoproduction on the nucleon, encounter 
various problems\cite{WiA93} (see Fig.~\ref{fig1}):
\begin{figure}[h] 
\centerline{\psfig{file=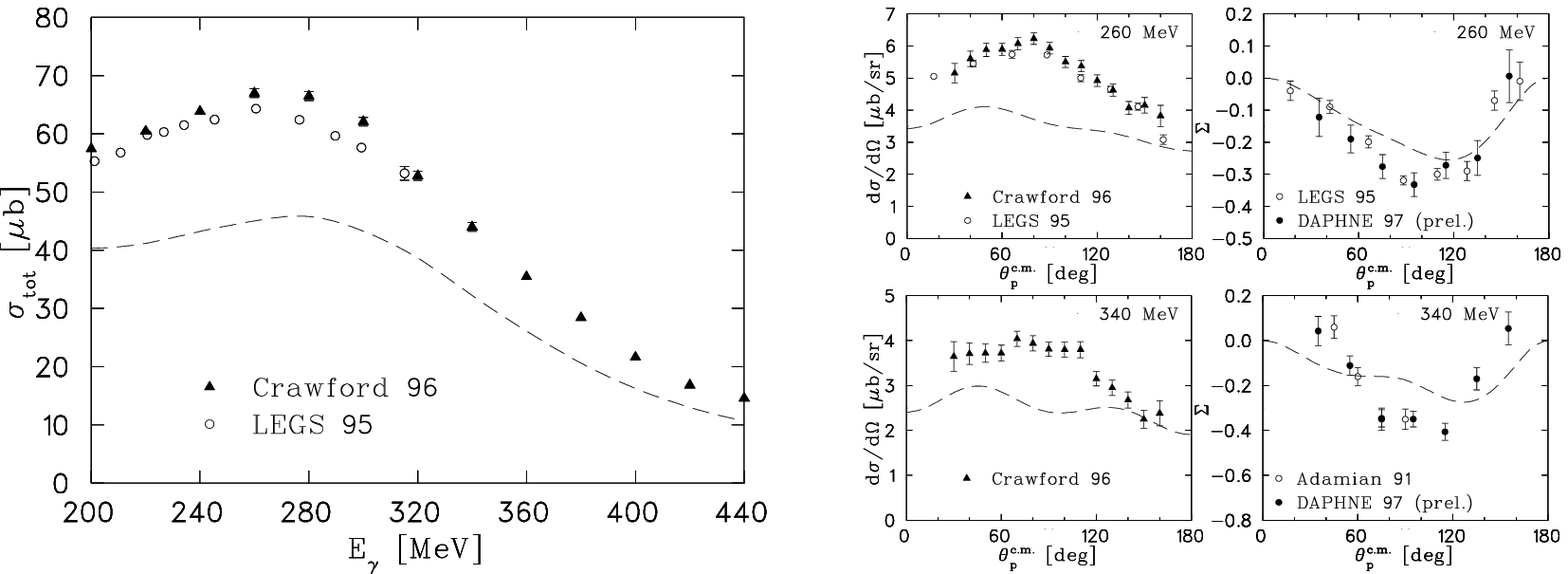,width=12cm}} 
\fcaption{Total and differential cross sections, and photon asymmetry 
of deuteron photodisintegration. 
Dashed: static calculation of Wilhelm et al.\protect\cite{WiA93}.
Exp.: R. Crawford {\it et al.}, \protect\Journal{\NPA}{603}{303}{1996},
LEGS 95: 
G. Blanpied {\it et al.}, \protect\Journal{\PRC}{52}{R455}{1995},
F. Adamian {\it et al.}, \protect\Journal{\JPG}{17}{1189}{1991},
DAPHNE 97:
S. Wartenberg {\it et al.}, \protect\Journal{\FBS}{26}{213}{1999}.
}
\vspace*{-.3cm}
\label{fig1}
\end{figure}
\begin{figure}[h] 
\centerline{\psfig{file=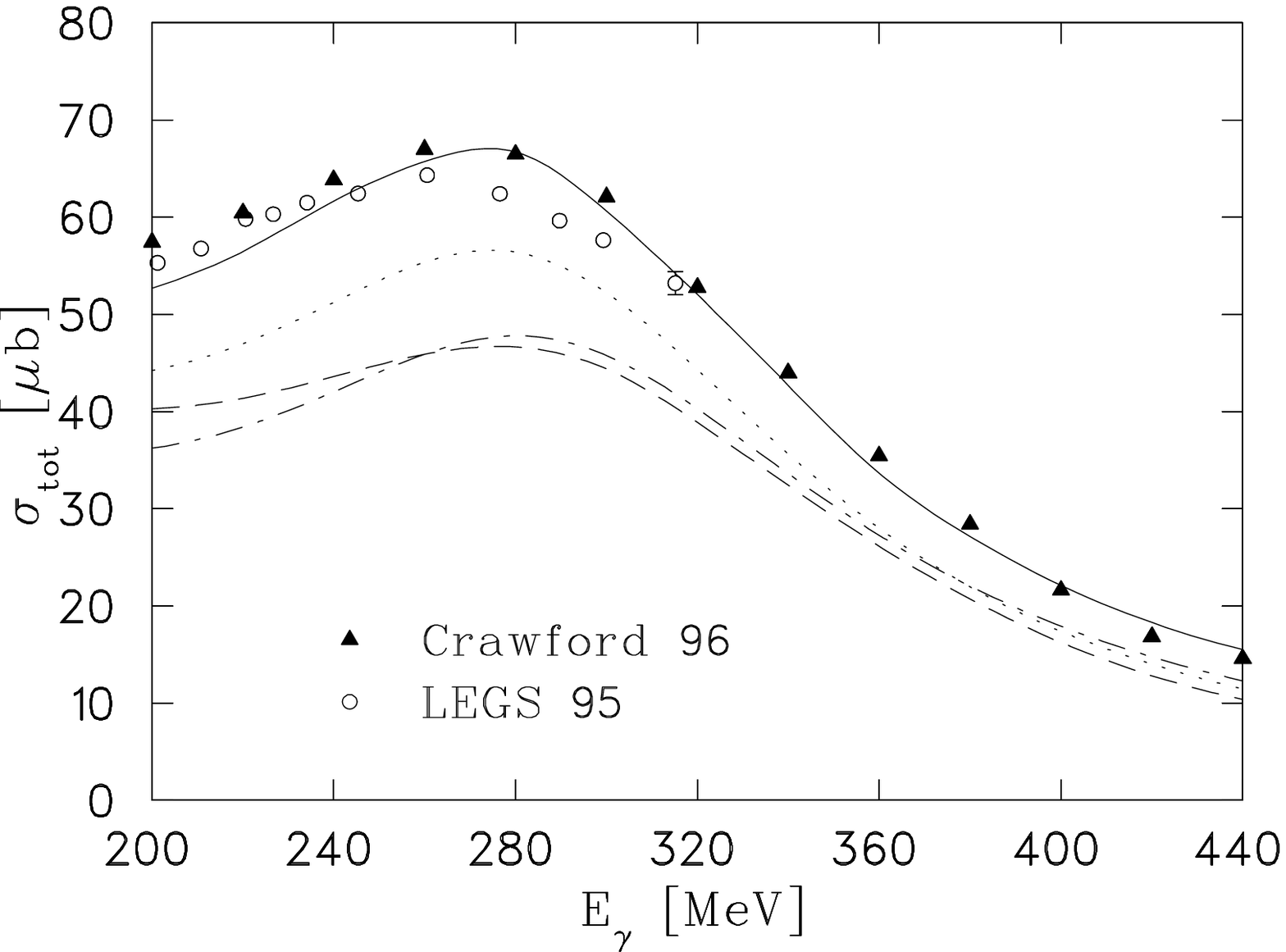,width=6cm}} 
\fcaption{Total cross section for $d(\gamma,p)n$. 
Dashed: static calculation of  
Wilhelm et al.\protect\cite{WiA93}; all other curves from\protect\cite{ScA98}: 
dotted: improved static calculation using Bonn OBEPR potential;
dash-dot: retardation in hadronic part but static MEC; 
solid: complete calculation including  $\pi d$-channel
and  $\rho \pi \gamma / \omega \pi \gamma$-MECs. Exp.\ as in
Fig.~\ref{fig1}.}
\label{fig3}
\end{figure}
\begin{figure}[h] 
\centerline{\psfig{file=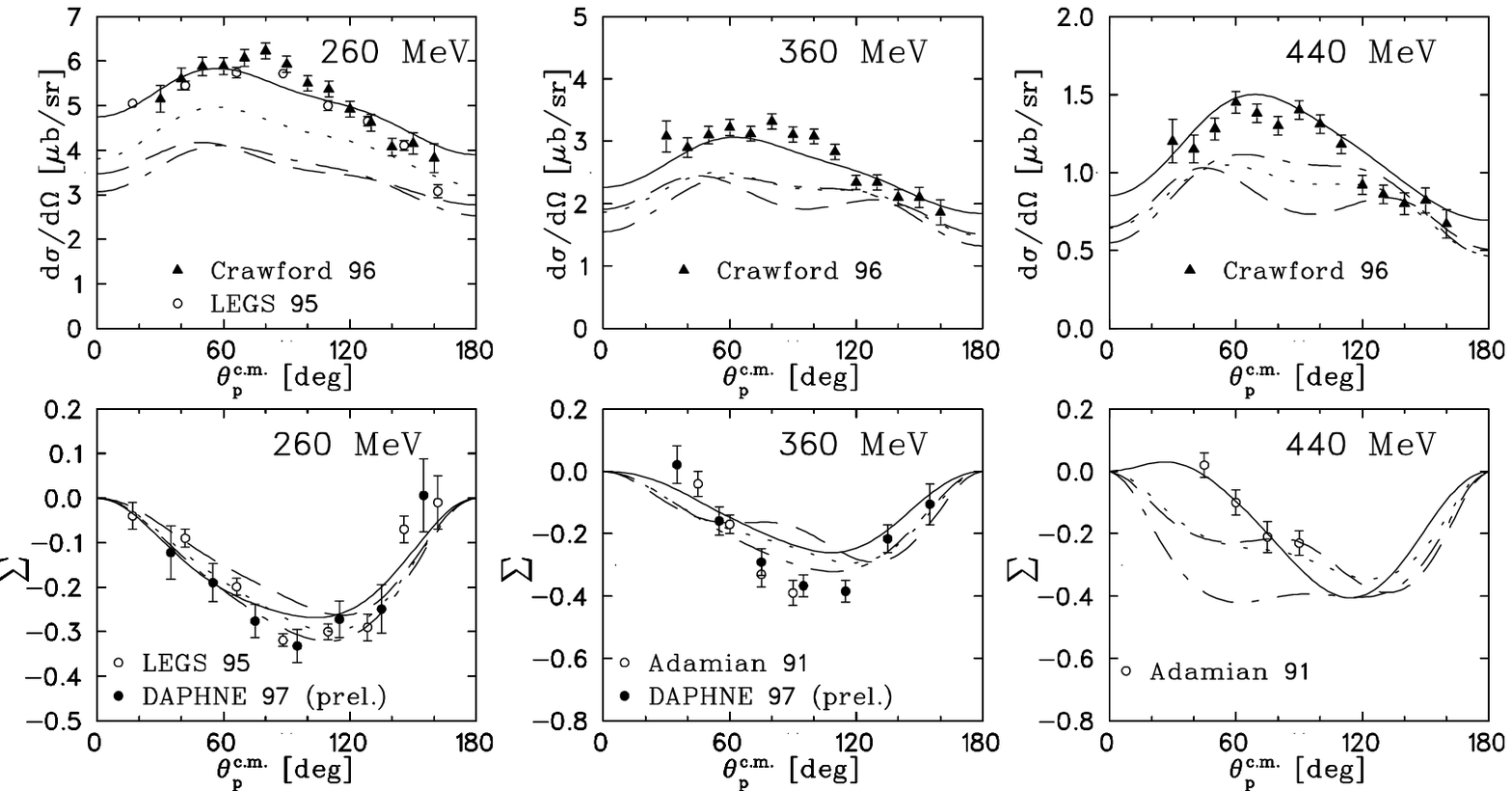,width=12cm}} 
\fcaption{Differential cross sections (upper panels) and photon
asymmetries (lower panels) of deuteron 
photodisintegration from\protect\cite{ScA98}. Notation as in
Fig.~\ref{fig3}. Exp. as in Fig.~\ref{fig1}.}
\vspace*{-6pt}
\label{fig4}
\end{figure}
(i)~Underestimation of the total cross section by 20-30\%.
(ii) Angular distributions develop a dip around 90 degree, especially 
above 300 MeV. (iii) The shape of the photon asymmetry is at variance 
with experimental data.
A detailed analysis of the role of the Born terms and their 
correspondence to MEC has 
lead to the conjecture, that $\pi$ retardation in potential and
exchange current might be important. Indeed, in a recent
calculation\cite{ScA98,ScA01} the importance of retardation in both,
the $NN$-interaction as well as MEC has been shown. 
For the total cross section this
is demonstrated in Fig.~\ref{fig3} where a quantitative agreement with
experimental data is achieved. Also for the differential cross section 
and the photon asymmetry shown in Fig.~\ref{fig4} one notes a much 
improved description. In particular, the dip in the angular distribution
has disappeared completely. Thus any realistic description of e.m.\ 
reactions on light nuclei in this energy regime has to use retarded
interactions and MEC.

\section{Off-shell Effects in One-Body Current}
\vspace*{-0.5pt}
\noindent
It is a well-known fact that the e.m.\ current of a particle with
internal structure becomes in general more complicated in an off-shell
situation. Consider, for example the e.m.\ current of a proton.
Its on-shell Dirac current is determined by two structure functions,  
the Dirac and Pauli form factors $F_1(q^2)$ and $F_2(q^2)$,
respectively, which depend on $q^2$, the four momentum transfer
squared, only. 
However, for e.m.\ reactions on nuclei the interacting nucleons are 
off-shell, i.e.\ $p^{\prime 2}=W^{\prime 2}\neq M^2 \neq W^2=p^2$. 
\begin{figure}[h] 
\centerline{\psfig{file=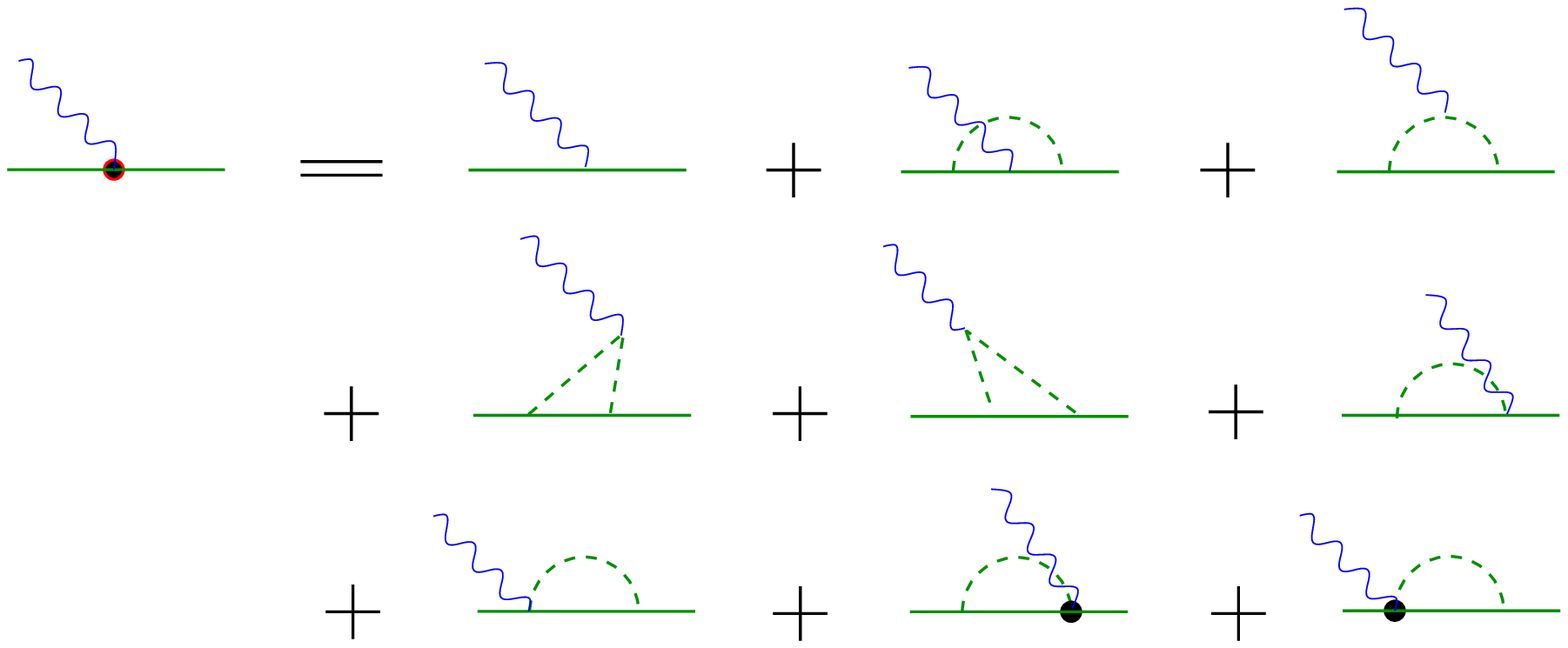,width=9cm}} 
\fcaption{Electromagnetic current of a nucleon dressed with a meson cloud.}
\vspace*{-8pt}
\label{fig8}
\end{figure}
In this case, 
one has ten additional structure functions depending not only
on $q^2$ alone but also on the off-shell masses $W'$ and $W$. 
The problem is that for the off-shell structure functions one needs a
dynamical model for the internal structure, because these form factors
are intimately connected to the underlying hadronic interaction. 
Thus, such a model has to be consistent with the $NN$-interaction.
Moreover, off-shell effects as such 
are not independently observable\cite{ScF01}. 
\begin{figure}[h] 
\centerline{\psfig{file=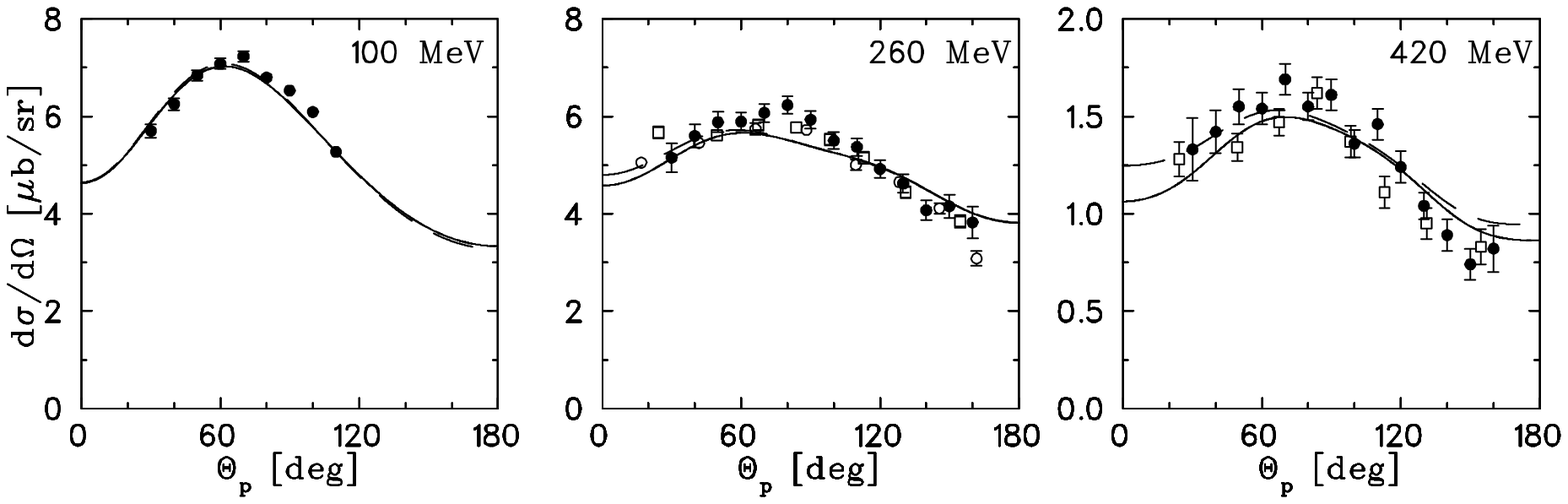,width=12cm}} 
\fcaption{Differential cross sections. Dashed curves without and solid 
with off-shell effects from\protect\cite{ScA01a}. 
Exp.: black circles: Crawford {\it et al.}, 
\protect\Journal{NPA}{603}{303}{1996}, open squares: Arends {\it et al.}, 
\protect\Journal{NPA}{412}{509}{1984}, open circles:
G. Blanpied {\it et al.}, \protect\Journal{\PRC}{52}{R455}{1995}.
}
\vspace*{-6pt}
\label{fig9}
\end{figure}
Recently, we have completed a study of off-shell effects in deuteron 
photodisintegration\cite{ScA01a} using a consistent dynamical approach  
in which the nucleon is dressed by pion loops, i.e.\ the internal 
nucleon structure is described by a pion cloud. 
Correspondingly, one finds for the e.m.\ one-nucleon current the 
representation depicted in Fig.~\ref{fig8}. 
For the resulting off-shell one-nucleon current one finds in the 
nonrelativistic limit
\begin{eqnarray*}
\langle \vec{p}^{\,\prime},e'|
\vec{j}(0)|\vec{p},e\rangle
\,&=&\, 2\,\beta(e',e)
\vec{p}^{\,\prime}+\gamma(e',e)\vec{k}
+\,i\vec{\sigma}\times \big(\delta(e',e)
\vec{p}^{\,\prime}+\epsilon(e',e)\vec{k}\big)\,,
\end{eqnarray*}
with $\vec{k}\,=\,\vec{p}^{\,\prime}-\vec{p}$, and $e'$ and $e$ denote
final and initial nucleon energies, respectively. The structure
functions fulfil the on-shell condition, i.e.\ for $e'=p^{\prime 2}/2M$
and $e=p^{2}/2M$ 
\begin{eqnarray*}
\beta(e',e)&\,=\,&-\gamma(e',e)=\frac{1}{2M}\,,\quad\delta(e',e)\,=\,0\,,
\quad\epsilon(e',e)\,=\,\frac{\mu}{2M}\,.
\end{eqnarray*}
The correct on-shell current is ensured by an appropriate counter
term. Within this meson-nucleon model, which is also used for the
$NN$-interaction including retardation effects,
one finds a sizeable influence from off-shell effects on the 
differential cross section of deuteron photodisintegration in and 
above the region of the $\Delta$-resonance as is seen in Fig.~\ref{fig9}. 
They show up predominantly at forward and backward angles, leading to
a decrease of the cross section. In the future, more realistic nucleon
models should be studied with respect to such off-shell effects. 

\section{Final State Interaction in Incoherent Pion Photoproduction}
\begin{figure}[h] 
\vspace*{6pt}
\centerline{\psfig{file=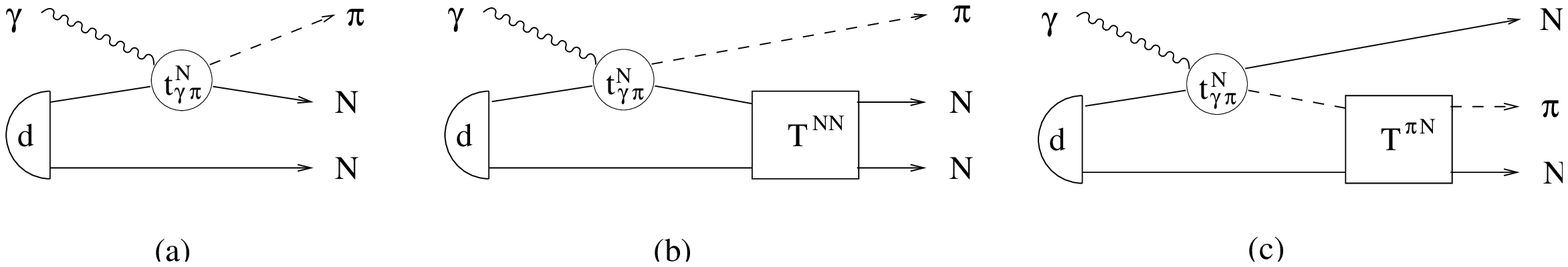,width=11cm}} 
\fcaption{Diagrams for incoherent pion photoproduction on the deuteron: 
(a) impulse approximation (IA), (b) $NN$ rescattering,
(c) $\pi N$ rescattering.}
\label{fig10}
\end{figure}

\vspace*{-0.5pt}
\noindent
Meson photoproduction on the nucleon provides valuable information on
its internal structure and serves as a test of hadron models.
The production on light nuclei is of particular interest because it
allows one to study the elementary neutron amplitude, medium effects
and nuclear structure. 
Recently, we have completed a calculation of pion photoproduction 
on the deuteron\cite{DaA02} including besides the impulse 
approximation complete rescattering in all two-body subsystems as 
depicted in Fig.~\ref{fig10}. 
\begin{figure}[h] 
\vspace*{-6pt}
\centerline{\psfig{file=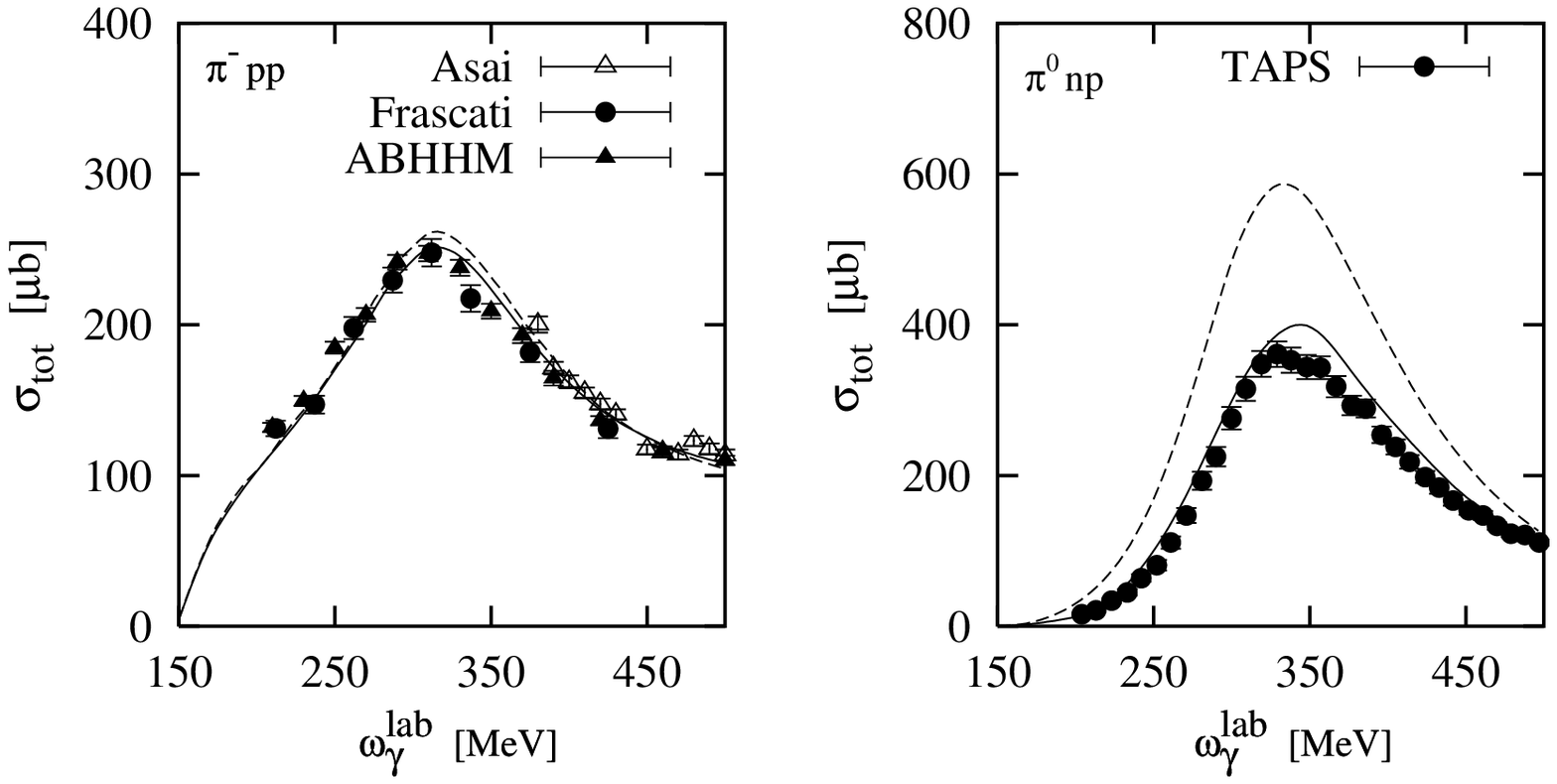,width=9cm}} 
\fcaption{Total cross section for pion photoproduction on the deuteron 
from\protect\cite{DaA02}. 
Dashed: IA; solid: IA+complete rescattering; data: Asai 
{\it et al.}, \Journal{\PRC}{42}{837}{1990}, Benz {\it et al.}, 
\Journal{\NPA}{65}{158}{1973} (ABHHM), Chiefari {\it et al.},  
\Journal{\LNC}{13}{129}{1975} (Frascati).}
\label{fig12}
\centerline{\psfig{file=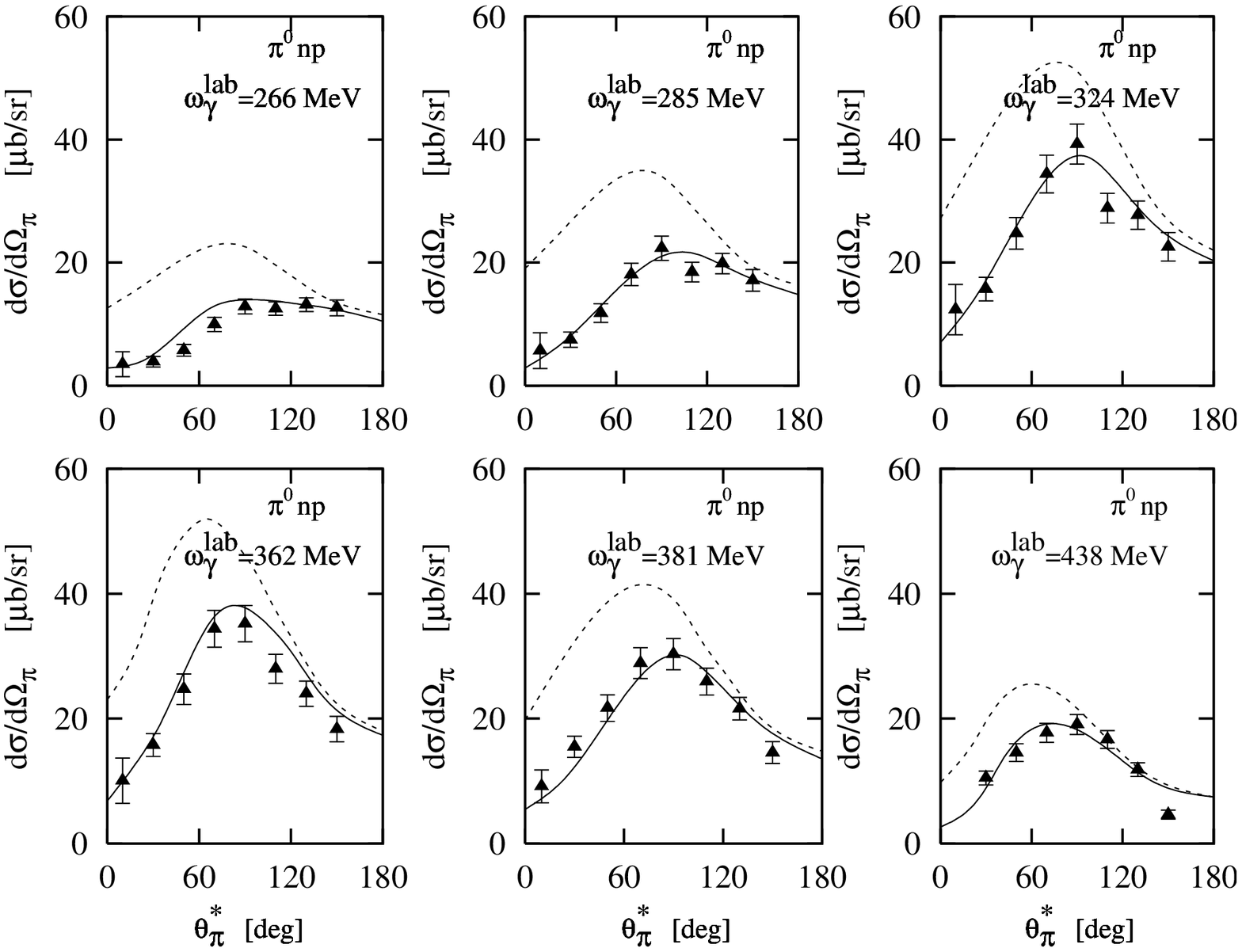,width=9cm}} 
\fcaption{Differential cross sections for 
\protect$d(\gamma,\pi^0)pn$ from\protect\cite{DaA02}. 
Dashed: IA; solid: IA+complete rescattering; 
Exp.: Krusche {\it et al.}, \Journal{\EPJA}{6}{309}{1999}.}
\vspace*{-6pt}
\label{fig13}
\end{figure}
Deuteron wave function and $NN$ interaction are taken from the Paris
potential, and the $NN$ and $\pi N$ interactions in separable form.
The complete $T$-matrices are obtained from solutions of the
corresponding LS-equations. For $NN$ rescattering
all partial waves with $J\leq 3$, and for $\pi N$ rescattering $S$
through $D$ waves are included. Total cross sections are shown in
Fig.~\ref{fig12}. For charged pion production the rescattering effect
is small, but 
for neutral pion production quite sizeable, which mainly stems from
the fact that in IA quite a fraction of the coherent production is
included due to the non-orthogonality of the final plane wave to the
deuteron bound state. In all cases $\pi N$-rescattering is very small compared
to $NN$-rescattering due to the much weaker $\pi N$-interaction. 
The inclusion of such rescattering contributions leads to a
satisfactory description of the experimental total cross sections 
for $\pi^-$ as well as for $\pi^0$ production as shown in 
Fig.~\ref{fig12}. A corresponding good agreement is achieved for 
the differential cross sections of $\pi^0$ production depicted 
in Fig.~\ref{fig13}. One readily notices that the major rescattering
effects appear at forward meson angles. This is also true for charged
pion production\cite{DaA02}. 

\section{Two-Body Effects in Coherent Eta Photoproduction}
\vspace*{-0.5pt}
\noindent
\begin{figure}[h] 
\centerline{\psfig{file=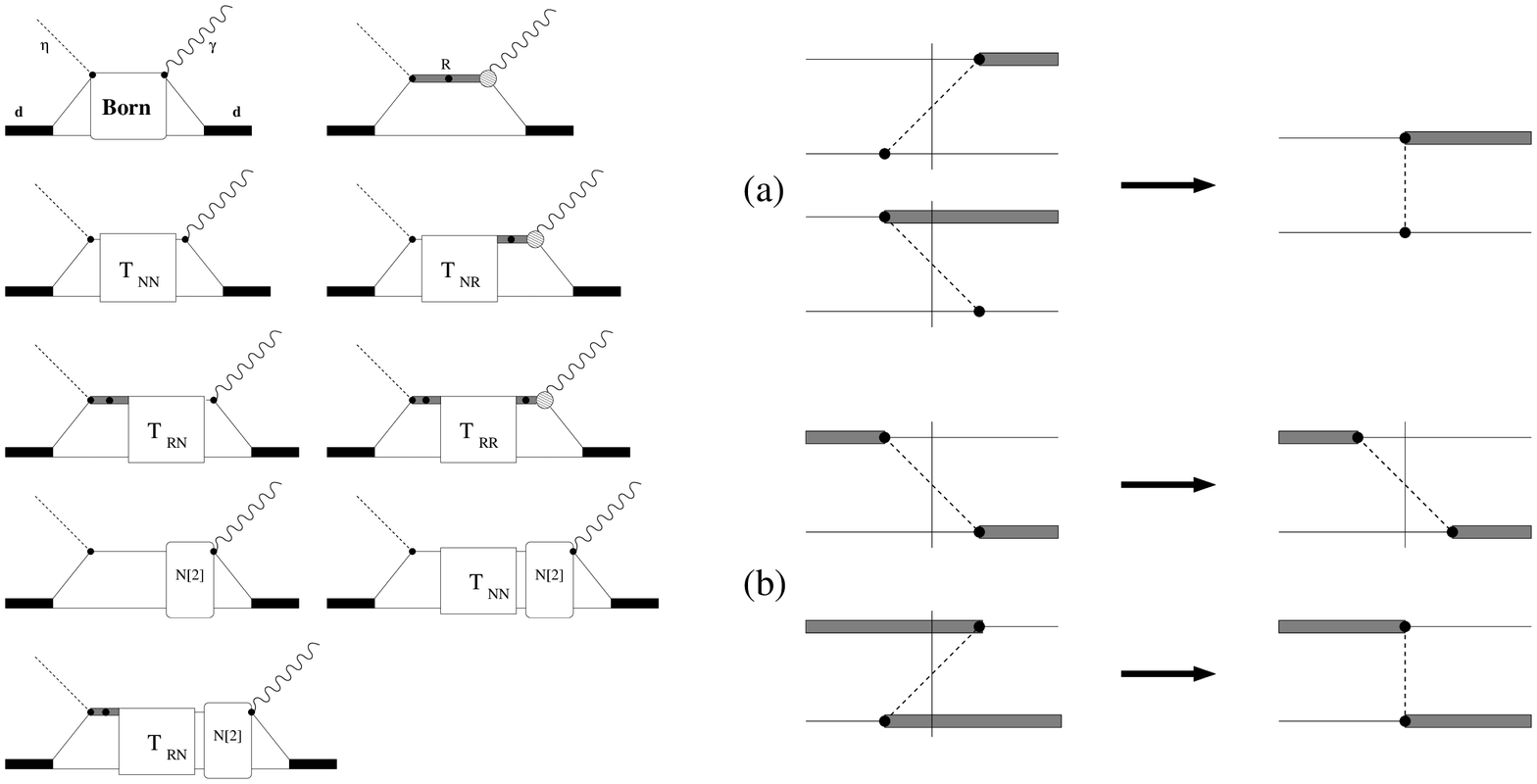,width=10cm}} 
\vspace*{6pt}
\fcaption{Left panel: Various mechanisms for $\eta$ production on the
deuteron. Right panel: Treatment of different time orderings of  
hadronic transition potentials. 
(a) static approximation for $NN\leftrightarrow{}NR$ potential,
(b) upper part: retarded $NR$ exchange potential,
 lower part: static approximation of the meson-RR propagator.
}
\label{fig16}
\vspace*{6pt}
\centerline{\psfig{file=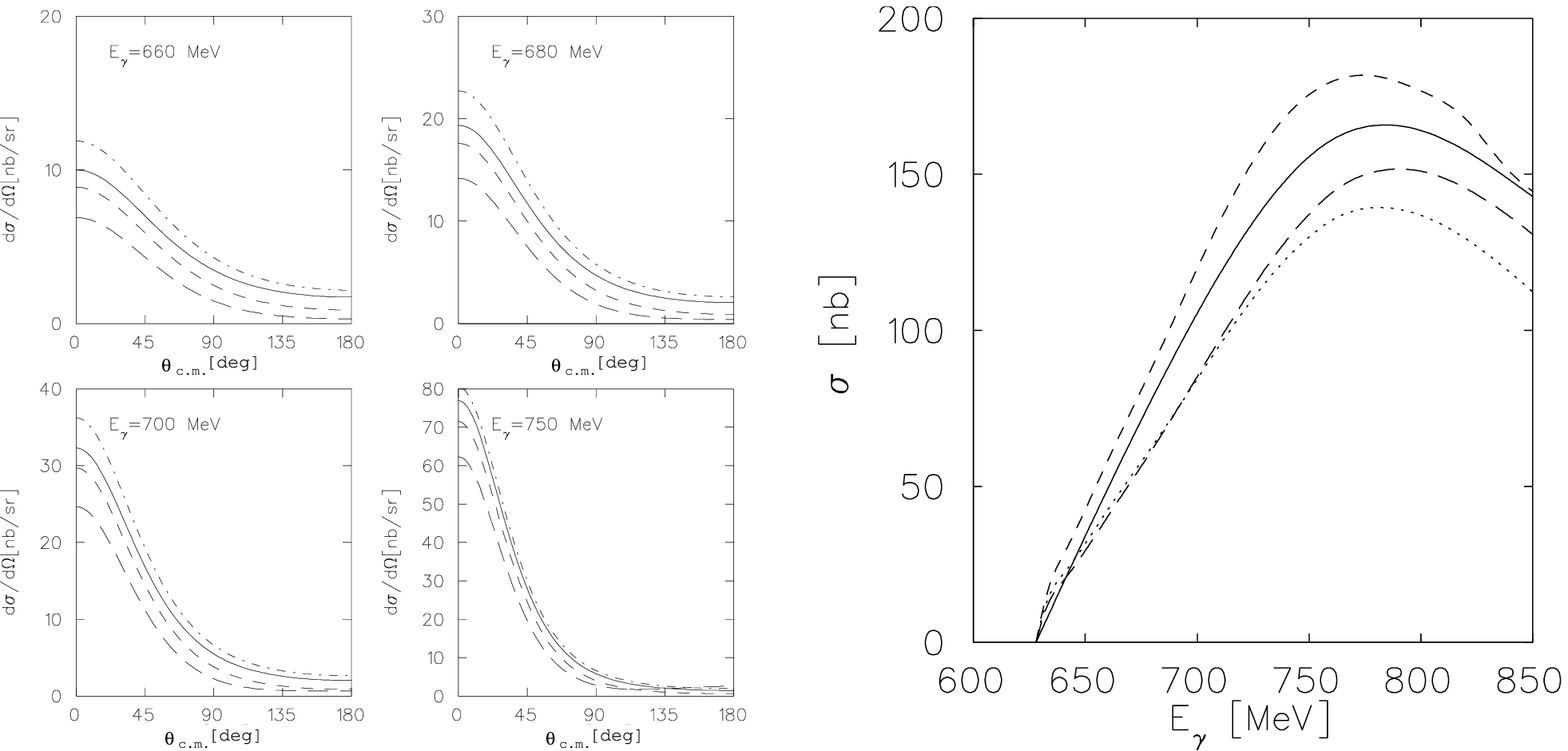,width=11cm}} 
\fcaption{Differential and total cross section for coherent 
$\eta$-production on deuteron. Left panel: short-dashed: IA, long-dashed: 
IA + $NN$ resc.\ + static $NS_{11}$-resc.; dash-dotted: 
IA + $NN$ resc.\ + retarded $NS_{11}$-resc.; 
solid: IA + all retarded rescatterings but no MEC. 
Right panel: dotted: pure resonance contribution; 
long-dashed: IA, short-dashed: IA + retarded first order rescattering,
solid: complete calculation.
}
\vspace*{-6pt}
\label{fig17}
\end{figure}
Photoproduction of $\eta$ mesons is an interesting tool for studying
the $(s=1/2)$-nucleon resonances. The $S_{11}(1535)$ plays a special role
because of its strong coupling to the $\eta$ channel. While the incoherent
reactions yields estimates of the modulus of the neutron amplitude, 
the coherent process provides the phase information. 
Furthermore, the deuteron with $T=0$ serves as isospin
filter and thus yields the ratio of isoscalar to proton amplitude
$A_s/A_p$. Until recently, a seeming discrepancy was noted between
extraction from the coherent reaction, $|A_s/A_p|_{coh}\approx 0.2$, and from
the incoherent one, $|A_s/A_p|_{incoh}=0.09$. The latter value was
extracted from the incoherent production on the deuteron, yielding
at resonance $(\sigma_n/\sigma_p)_{res}=0.66$. 

In recent work on coherent photoproduction on the deuteron\cite{RiA01}
we have 
taken for the elementary production amplitude a coupled channel model
of\cite{BeT91}, which considers the channels 
$\pi N\rightarrow \pi N$, $\pi N\rightarrow \eta N$, 
$\gamma N\rightarrow \pi N$, and $\gamma N\rightarrow \eta N$. 
Specific features of the model are the inclusion of 
self energy contributions from $\pi$ and $\eta$ loops and the 
dressing of the e.m.\ vertex by hadronic rescattering.
\begin{figure}[h] 
\centerline{\psfig{file=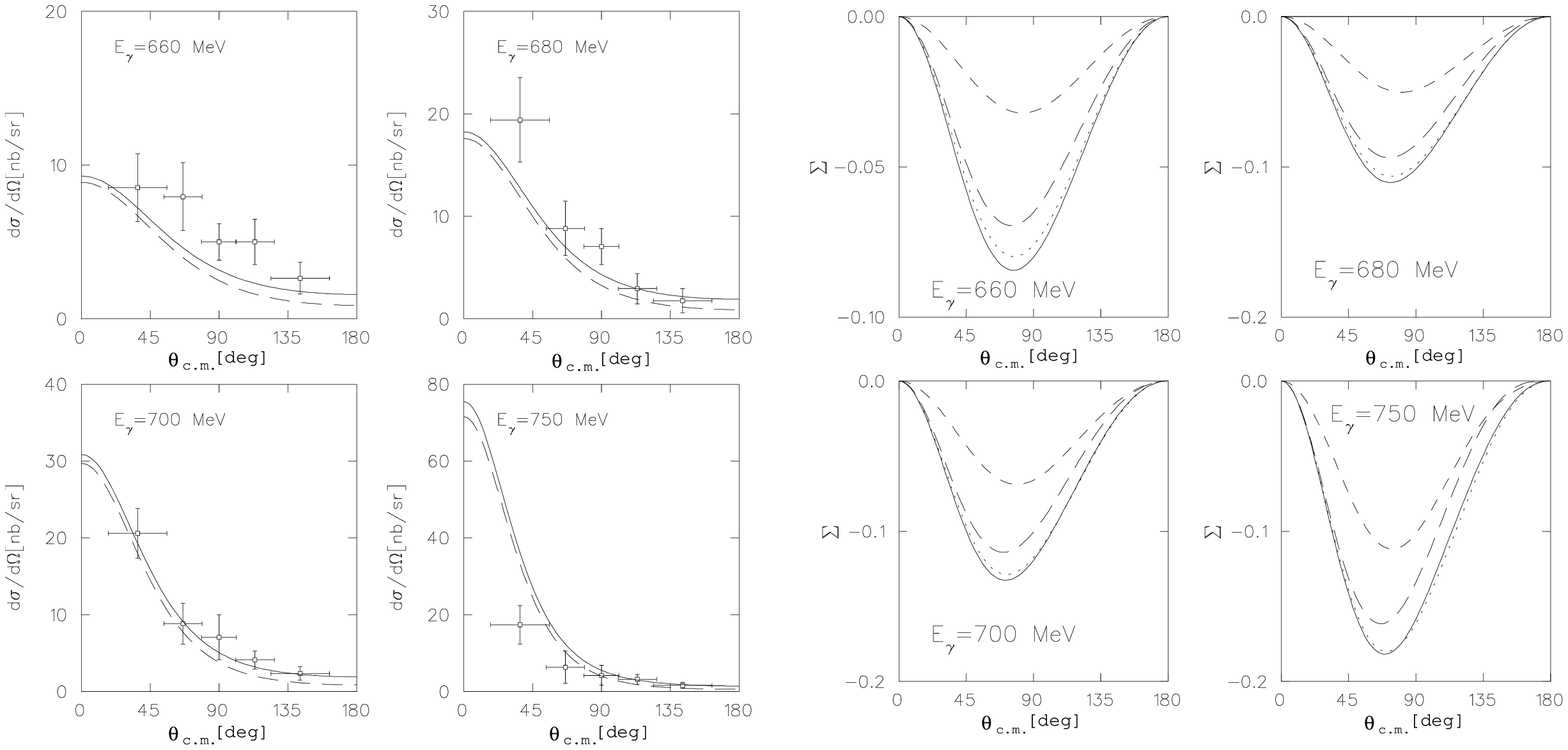,width=12cm}} 
\fcaption{Left panel: Differential cross section: dashed: IA,
solid: complete calculation; data: Hoffmann-Rothe {\it et al.}, 
\protect\Journal{\PRL}{78}{4697}{1997}. Right panel: Linear photon asymmetry: 
short-dashed: pure resonant contribution, 
long-dashed: IA, dotted: IA + retarded rescattering,
solid: complete calculation.
}
\vspace*{-6pt}
\label{fig18}
\end{figure}
A good description of $\eta$ photoproduction on the proton is achieved
with the important result that the dressing leads to complex values
for proton and neutron amplitudes, i.e., $A_n = ( -114-i1.7 )\times
10^{-3} \mbox{GeV}^{-1/2}$, $A_p =  ( 120.9-i66.1 )\times 10^{-3}
\mbox{GeV}^{-1/2}$, yielding the ratios 
$(\sigma_n/\sigma_p)_{res} = |A_n/A_p|^2 = 0.68 \approx 2/3$, and 
$A_s/A_p = 0.25\,e^{-i 0.969}$. Thus there is no contradiction for
this ratio anymore between the extraction of this ratio from the
coherent and the incoherent reaction. 

In the left panel of Fig.~\ref{fig16} the various mechanisms, taken 
into account in\cite{RiA01}, are displayed. The box labeled Born 
contains disconnected diagrams where the photon is 
absorbed by one nucleon and the $\eta$ is emitted by the other.
Hadronic rescattering is indicated by boxes T$_{\mathrm {NN}}$, 
T$_{\mathrm {NR}}$, T$_{\mathrm {RN}}$, and T$_{\mathrm {RR}}$ and 
meson exchange current contributions by boxes N[2]. As resonances 
``R'' we have included $P_{11}(1440)$, $S_{11}(1535)$ and $D_{13}(1520)$. 
The hadronic interaction is considered in static as well as retarded 
form as displayed in the right panel of Fig.~\ref{fig16}. 
For coherent $\eta$ photoproduction on the deuteron, the effect of
various mechanisms on the differential cross section are displayed
in the left panel of Fig.~\ref{fig17}. One notes an opposite effect 
between static and retarded rescattering. The total cross section 
is shown in the right panel of Fig.~\ref{fig17}, where a sizeable 
influence from all two-body effects is seen. Furthermore, the first
order rescattering approximation overestimates considerably the 
total cross section.
For the differential cross section a reasonable though not perfect 
agreement with experiment is achieved as is shown in the left panel of 
Fig.~\ref{fig18}. In the right panel of this figure the photon asymmetry 
is displayed. One notes a sizeable influence from hadronic rescattering
but little effect from MEC.

\section{Final State Interaction in Incoherent Eta Photoproduction}
\vspace*{-0.5pt}
\noindent
\begin{figure}[h] 
\centerline{\psfig{file=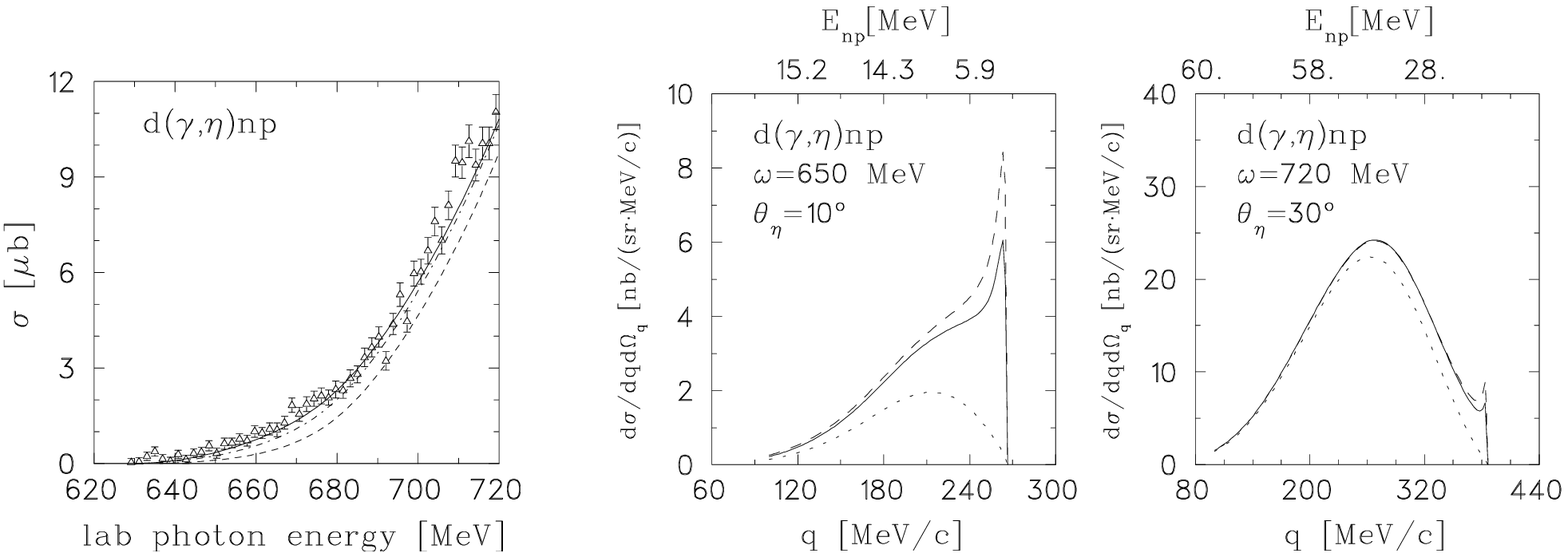,width=12cm}} 
\fcaption{Left panel: Total cross section for $d(\gamma,\eta)np$
from\protect\cite{FiA97}: dashed: IA, 
solid: IA + rescattering, dash-dotted: IA + $NN$ rescattering,
data: inclusive $\gamma d\!\rightarrow\eta X$ from\protect\cite{Kru95}.
Middel and right panels: $\eta$-meson spectra at forward emission angles
for two different photon energies and angles: dotted: IA,
solid: IA + $NN$ rescattering, dashed: without the deuteron 
$D$-wave in the $NN$-rescattering contribution. 
The excitation energy $E_{np}$ in the final
$NN$-system is indicated at the top abscissa.
}
\label{fig20}
\vspace*{6pt}
\centerline{\psfig{file=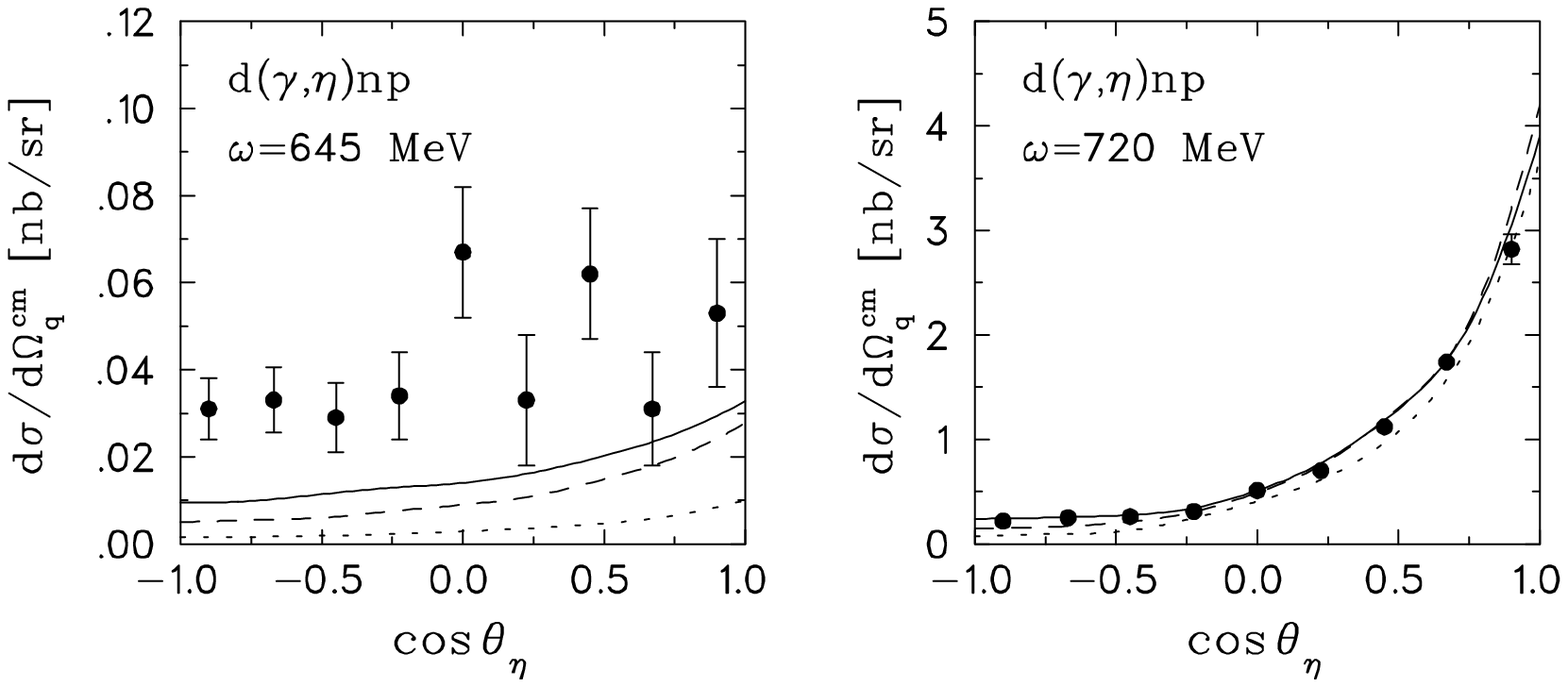,width=10cm}} 
\fcaption{Differential cross sections near threshold from\protect\cite{FiA97}.
Dotted: IA, dashed: IA + $NN$ rescattering;
solid: IA + complete rescattering. Exp.\ from Krusche {\it et al.},
\protect\Journal{\PLB}{358}{40}{1995}.
}
\label{fig21}
\vspace*{-.5cm}
\end{figure}
Near threshold the impulse approximation yields a very small cross 
section for $d(\gamma,\eta)np$ due to the large momentum mismatch 
and indeed fails drastically compared to experiment yielding 
a cross section much too low.\\ 
\centerline{\psfig{file=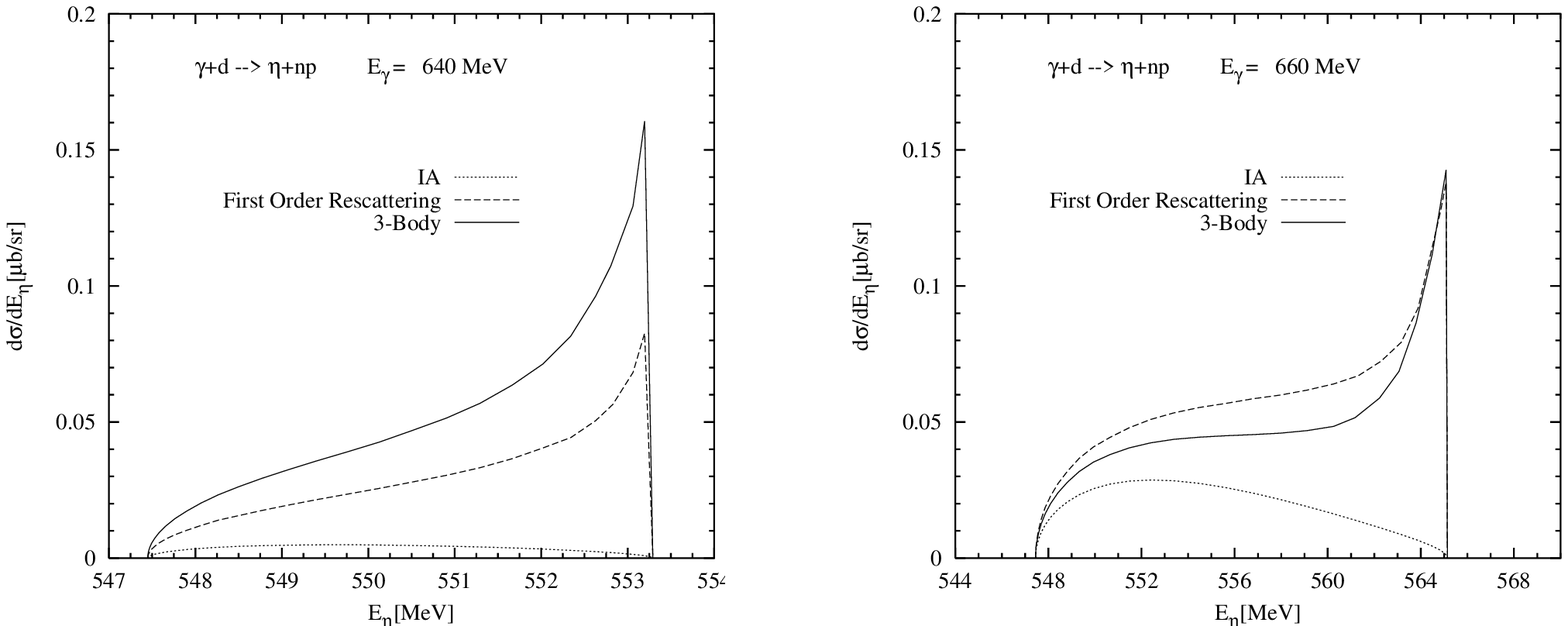,width=10cm}} 
\fcaption{$\eta$ meson spectrum for $d(\gamma,\eta)np$. Dotted: IA, dashed: 
first order rescattering, solid: complete thre-body calculation.}
\label{fig24a}
Therefore, we
first have performed an approximate treatment of FSI\cite{FiA97} in complete
analogy to pion photoproduction on the deuteron, i.e.\ taking into 
account only complete rescattering in the two-body $\eta N$ and $NN$ 
subsystems in the final state. In the following, this is called 
first order rescattering. 
In this case the $NN$ $t$-matrix is determined from the Bonn OBEPQ
potential and for the $\eta N$ $t$-matrix an isobar model is taken 
describing the intermediate $S_{11}(1535)$ excitation. 
The first order rescattering, restricted to $s$-waves of $NN$ and
$\eta N$ subsystems in view of the near-threshold region, 
leads to a considerable improvement (see left panel of
Fig.~\ref{fig20}). 
The spectrum of the outgoing $\eta$ meson (middel and right panels of
Fig.~\ref{fig20}) shows the distinct signature
of the final state $NN$ rescattering exhibiting the prominent $^1S_0$ 
peak near the $NN$ scattering threshold.
\begin{figure}[h] 
\centerline{\psfig{file=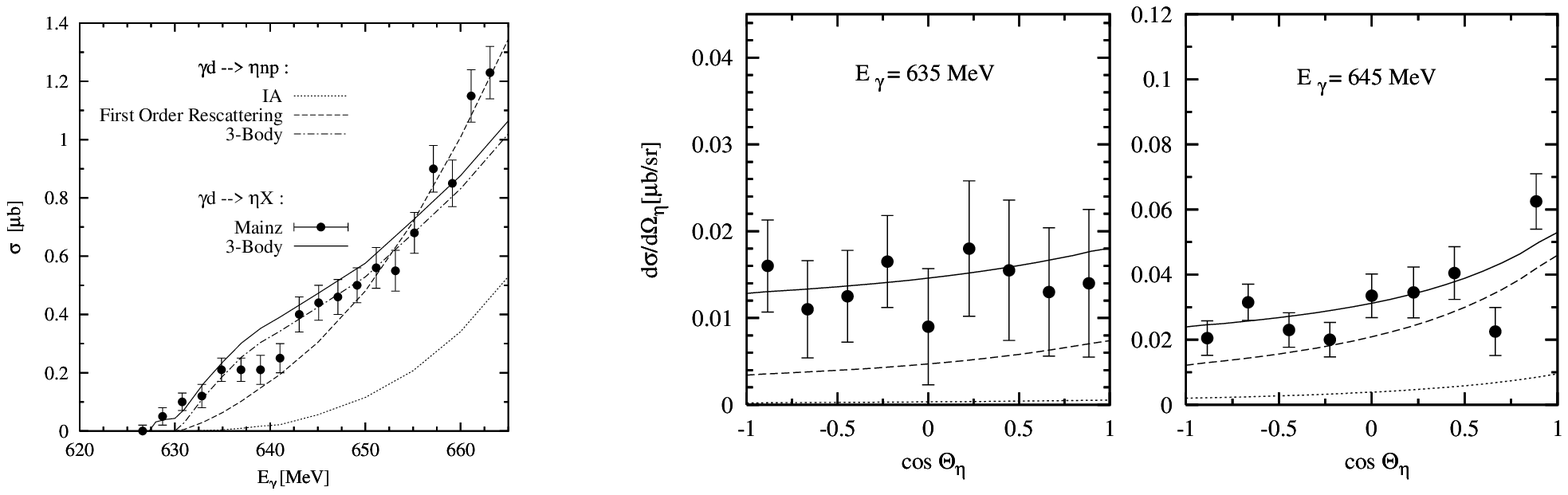,width=12cm}} 
\fcaption{Results for $d(\gamma,\eta)np$ from\protect\cite{FiA02}. Left
panel: Total cross section: 
(a) incoherent: dotted: IA; dashed: first order rescattering;
dash-dot: complete 3-body model; (b) inclusive:
solid: sum of coherent and incoherent channels.
Inclusive data (Mainz) from Hejny {\it et al.}\protect\cite{Hej02}. 
Middel and right panels: Differential cross sections: dotted: IA; 
dashed: first order rescattering; solid: complete 3-body model.
Inclusive data from Hejny {\it et al.}\protect\cite{Hej02}. 
}
\vspace*{-6pt}
\label{fig24}
\end{figure}
The differential cross sections near threshold are shown in Fig.~\ref{fig21}.
The left panel of Fig.~\ref{fig21} indicates that first order 
rescattering still fails to explain quantitatively the enhancement
of the data right above threshold. 
This is corroborated by very recent more precise near-threshold data
of Hejny {\it et al.}\cite{Hej02}.

For this reason, we then have performed a three-body treatment of 
the final state interaction\cite{FiA02}, 
because the very strong effect in first order rescattering suggests
that a genuine three-body treatment is required. 
A considerable simplification is achieved by restriction to only $s$-waves 
which is justified because of threshold region. 
For the $NN$ interaction a simple Yamaguchi form is used. The 
resulting $\eta$ spectrum is displayed in Fig.~\ref{fig24a} where
again one notes clearly the $^1S_0$ peak as in Fig.~\ref{fig20}. 
However, for the lower photon energy one readily sees a substantial 
underestimation of the first order rescattering compared to the 
three-body calculation, althoug the forms are similar. 
Total and differential cross sections are shown in Fig.~\ref{fig24}.
The inclusive total cross section data exhibit a distinct enhancement
near threshold which is reproduced by the 3-body approach (left panel
of Fig.~\ref{fig24}) but not in first order, the latter being considerably 
lower right at threshold. This is also the case for
the differential cross sections (middel and right panels of
Fig.~\ref{fig24}). It remains to be seen, whether a more realistic
treatment of the $NN$-interaction is also able to describe the data. 

\section{Conclusions and outlook}
\vspace*{-0.5pt}
\noindent
In summary, we may conclude that the electromagnetic probe is a very
important tool in order to reveal the internal structure of hadrons.
Only the new generation of high duty cycle machines allows one to 
exploit its full power and the thrust of future experimental research 
lies on the study of exclusive reactions. Polarization observables
will give us much more detailed information and thus will provide much
more stringent tests for theoretical models. 

Reactions on the deuteron
are of particular importance for testing present theoretical
frameworks for describing strong interaction physics in terms of
effective degrees of freedom, thus serving as a test laboratory. 
Of special interest are e.m.\ reactions above the pion threshold.
An important example is photodisintegration with respect to the study
of retardation and off-shell effects. 

Furthermore, meson production on
the deuteron offers the possibility to study the elementary production
amplitude on the neutron provided one has the two-body effects from final
state interaction and e.m.\ current under control. In particular, the
$\eta N$-interaction can be studied in incoherent eta production near
threshold. However, a first order rescattering 
calculation as used, e.g. in\cite{SiS02}, is not reliable for that purpose, 
because right above threshold a three-body approach is mandatory. 

Finally, for increased energy and momentum transfers, the effects
which arise from relativity should be carefully considered.


\begin{thebibliography}{99}

\bibitem{WiA93}
P. Wilhelm and H. Arenh\"ovel, \Journal{\PLB}{318}{410}{1993}.

\bibitem{ScA98}
M. Schwamb, H. Arenh\"ovel, and P. Wilhelm, \Journal{\PLB}{420}{255}{1998}.

\bibitem{ScA01}
M. Schwamb and H. Arenh\"ovel, \Journal{\NPA}{690}{682}{2001}.

\bibitem{ScF01}
S. Scherer and H. Fearing, \Journal{\NPA}{684}{499}{2001}.

\bibitem{ScA01a}
M. Schwamb and H. Arenh\"ovel, \Journal{\NPA}{696}{556}{2001}.

\bibitem{DaA02}
E.M. Darwish, H. Arenh\"ovel, and M. Schwamb, nucl-th/0208030; 
{\it Eur.\ Phys.\ J.} A (in print).

\bibitem{RiA01}
F. Ritz and H. Arenh\"ovel, \Journal{\PRC}{64}{034005}{2001}.

\bibitem{BeT91}
C. Bennhold and H. Tanabe, \Journal{\NPA}{530}{625}{1991}.

\bibitem{FiA97}
A. Fix and H. Arenh\"ovel, \Journal{\ZPA}{359}{427}{1997}.

\bibitem{FiA02}
A.\ Fix and H. Arenh\"ovel, \Journal{\NPA}{697}{277}{2002}.

\bibitem{Kru95}
B. Krusche {\it et al.}, \Journal{\PLB}{358}{40}{1995}.

\bibitem{Hej02}
V. Hejny {\it et al.}, \Journal{\EPJA}{13}{493}{2002}.

\bibitem{SiS02}
A. Sibirtsev {\it et al.}, \Journal{\PRC}{65}{044007}{2002}.

\end{thebibliography}
\end{document}